\documentclass[11pt]{article}
\usepackage{graphicx}
\usepackage{longtable,lscape}
\usepackage{amsfonts}
\usepackage{bbm}
\usepackage{float}
\usepackage{url}

\newcommand{\captionfonts}{\footnotesize}
\makeatletter  % Allow the use of @ in command names
\long\def\@makecaption#1#2{%
  \vskip\abovecaptionskip
  \sbox\@tempboxa{{\captionfonts #1: #2}}%
  \ifdim \wd\@tempboxa >\hsize
    {\captionfonts #1: #2\par}
  \else
    \hbox to\hsize{\hfil\box\@tempboxa\hfil}%
  \fi
  \vskip\belowcaptionskip}
\makeatother 
\begin{document}
\title{A Quantum Model for the Ellsberg and \\ Machina Paradoxes}
\author{Diederik Aerts and Sandro Sozzo \vspace{0.5 cm} \\ 
        \normalsize\itshape
        Center Leo Apostel for Interdisciplinary Studies \\
        \normalsize\itshape
        Brussels Free University \\ 
        \normalsize\itshape
         Krijgskundestraat 33, 1160 Brussels, Belgium \\
        \normalsize
        E-Mails: \url{diraerts@vub.ac.be,ssozzo@vub.ac.be} \\ \\
        Jocelyn Tapia \vspace{0.5 cm} \\ 
        \normalsize\itshape
        Pontificia Universidad Cat\'olica de Chile \\
        \normalsize\itshape
         Avda. Libertador Bernardo OHiggins 340, Santiago, Chile \\
        \normalsize
        E-Mail: \url{jntapia@uc.cl}
        }
\date{}
\maketitle
\begin{abstract}
\noindent
The \emph{Ellsberg} and \emph{Machina paradoxes} reveal that \emph{expected utility theory} is problematical when real subjects take decisions under uncertainty. Suitable generalizations of expected utility exist which attempt to solve the Ellsberg paradox, but none of them provides a satisfactory solution of the Machina paradox. In this paper we elaborate a quantum model in Hilbert space describing the Ellsberg situation and also the Machina situation, and show that we can model the specific aspect of the Machina situation that is unable to be modeled within the existing generalizations of expected utility.
\end{abstract}


\section{Introduction\label{intro}}
In economics, the predominant model of decision making is the {\it Expected Utility Theory} (EUT) \cite{vonneumannmorgenstern1944,savage1954}. Notwithstanding
its simplicity, mathematical tractability and predictive success, 
the empirical validity of EUT at the individual level is questionable. Indeed, examples exist in the literature which show an inconsistency between real preferences and the predictions of EUT. These deviations were put forward by considering specific situations of uncertainty often commonly referred to now as paradoxes \cite{ellsberg1961,machina2009}.

EUT was formally developed by von Neumann and Morgenstern \cite{vonneumannmorgenstern1944}. They presented a set of axioms that allow to represent decision--maker preferences over the set of \emph{acts} (functions from the set of states of the world into the set of consequences) by the functional $E_p u(.)$, for some real--valued Bernoulli utility function $u$ on the set of consequences and an objective probability measure $p$ on the set of states of the nature. An important aspect of EUT concerns the treatment of uncertainty. Knight had highlighted the difference between \emph{risk} and \emph{uncertainty} reserving the term \emph{risk} for ventures that can be described by known (or physical) probabilities, and the term \emph{uncertainty} to refer to situations in which agents did not know the probabilities associated with each of the possible outcomes of an act \cite{knight1921}. However, probabilities in the von Neumann and Morgenstern modeling are {\it objectively} or, physically, given. Later, Savage extended EUT allowing agents to construct their own subjective probabilities when physical probabilities are not available \cite{savage1954}. Then according to Savage's model, the distinction put forward by Knight seems to be irrelevant. Ellsberg's experiments instead showed that Knightian's distinction is empirically meaningful \cite{ellsberg1961}. In particular, he presented the following experiment. Consider one urn with thirty red balls and sixty balls that are either yellow or black, the latter in unknown proportion. One ball will be drawn from the urn. Then, free of charge, a person is asked to bet on one of the acts $f_1$, $f_2$, $f_3$ and $f_4$ defined in Table \ref{table01}.
\noindent
\begin{table} \label{table01}
\begin{center} 
\begin{tabular}{cccc}
\hline
Act & red & yellow & black\\
\hline
$f_1$ & 12\$ & 0\$ & 0\$   \\
$f_2$ & 0\$ & 0\$ & 12\$  \\
$f_3$ & 12\$ & 12\$ & 0\$  \\
$f_4$ & 0\$ & 12\$ & 12\$  \\
\hline
\end{tabular}
\end{center}
\caption{The payoff matrix for the Ellsberg paradox situation.}
\end{table}
\noindent
When asked to rank these gambles most of the persons choose to bet on $f_1$ over $f_2$ and $f_4$ over $f_3$. This  empirical result cannot be explained by EUT. In fact, we can see that individuals' ranking of the sub--acts [12 on {\it red}; 0 on {\it black}] versus [0 on {\it red}; 12 on {\it black}] depends upon whether the event $yellow$ yields a payoff of 0 or 12, contrary to what is suggested by the Sure--Thing principle, an important axiom of Savage's model. Nevertheless, these choices have a direct intuition: $f_1$  offers the 12 prize with an {\it objective probability} of $1/3$, and $f_2$ offers the same prize but in an element of the {\it subjective partition}  $\{black, yellow  \}$. In the same way, $f_4$ offers the prize with an objective probability of $2/3$, whereas $f_3$ offers the same payoff on the union of the unambiguous event {\it red} and the ambiguous event {\it yellow}. Thus, in both cases the unambiguous bet is preferred to its ambiguous counterpart, a phenomenon called by Ellsberg {\it ambiguity aversion}. 

After the work of Ellsberg many extensions of EUT have been developed to represent this kind of preferences, all replacing the Sure--Thing Principle by weaker axioms. The first extension is {\it Choquet Expected Utility} also known as expected utility with {\it non--additive} probabilities \cite{gilboa}. This model considered a subjective non--additive probability \emph{(capacity}) over the states of nature instead of a subjective probability.
Thus, decision--makers could underestimate or overestimate probabilities in the Ellsberg experiment and the ambiguity aversion is equivalent to the convexity of the capacity (pessimistic beliefs).
A second approach is the {\it $Max-Min$ Expected Utility}, or expected utility with multi--prior \cite{gilboaschmeidler1989}. In this case the lack of knowledge about the states of nature of the decision--maker cannot be represented by a unique probability measure, instead  he or she thinks are relevant  a set of probability measures, then an act $f$ is preferred to $g$ if $ \min_{p\in P} E_p u(f) >  \min_{p\in P} E_p u(f)$, where $P$ is a convex and closed set of additive probabilities measures. The ambiguity aversion is represented by the pessimistic beliefs of the agent which takes decisions considering the worst probabilistic scenario. 
The third model is {\it Variational Preferences} \cite{mmr2006}, and it is a dynamic generalization of the Max-Min expected utility. In this case agents rank acts according to the criterion: $\inf_{p\in \bigtriangleup } \{E_p u(f)+c(p)\}$, where $c(p)$ is a closed and convex penalty function associated with the probability election. 
Finally, the {\it Second Order Probabilities} approach \cite{kmm2005} proposes a model of preferences over acts such that the decision--maker prefers an act $f$ to an act $g$ if and only if $E_{\mu} \phi (E_p u (f) ) \geq  E_{\mu} \phi(E_p u (g))$, where E is the expectation operator, u is a von Neumann--Morgenstern utility function, $\phi $ is an increasing transformation, and $\mu $ is a subjective probability over the set of probability measures $p$ that the decision--maker thinks are feasible. In this kind of model the ambiguity aversion is represented by the concavity of the transformation $\phi$. 

Notwithstanding the models above have been widely used in economics and finance, they are not absent of critics (see, e.g., \cite{machina2009,epstein1999}). In the same spirit as Ellsberg, Machina proposed an example introducing a trade off between ambiguity aversion and Bayesian advantages that cannot be represented by the Choquet expected utility model \cite{machina2009}. 
Recently it was proved that no one of the 
mentioned extensions of EUT 
can represent the behavior described by the Machina paradox \cite{bdhp2011}. 
For the Machina paradox an experiment 
is considered consisting of an urn with four kind of different balls identified with a number between 1 and 4. The amount of balls with the number 1 plus the amount of balls with the number 2 is fifty and the amount of balls with the number 3 plus the amount of balls with the number 4 is fifty--one. 
Agents are asked to rank the following set of acts.
\noindent
\begin{table} \label{table02}
\begin{center}
\begin{tabular}{ccccc}
\hline
Act & $E_1$ & $E_2$ & $E_3$ & $E_4$\\
\hline
$f_1$ & 202 & 202 & 101 & 101  \\
$f_2$ & 202 & 101 & 202 & 101 \\
$f_3$ & 303 & 202 & 101 & 0  \\
$f_4$ & 303 & 101 & 202 & 0  \\
\hline
\end{tabular}
\end{center}
\caption{The payoff matrix for the Machina paradox situation.}
\end{table}
\noindent
The event $E_j$ indicates that a ball with a number $j$ has been drawn from the urn, the act $f_1$ has been defined as contingent payoff in each event, so that in $E_1$, $f_1$ pays 202, in $E_2$, $f_1$ pays 202, and so on. Equally are defined $f_2$, $f_3$ and $f_4$. Then, free of charge a person is asked to bet on $f_1$ or to bet on $f_2$, if he or she are sufficiently uncertainty averse then will prefer $f_1$ instead $f_2$, because $f_1$ has not ambiguity in its payoffs although $f_2$ presents a slight Bayesian advantage due to the 51 balls may yield 202. The person is also asked to bet on $f_3$ or $f_4$. In this case, both acts present ambiguity in their payoffs, there is not an informational advantage between them. Thus a decision--maker who values unambiguous information would be indifferent between $f_3$ and $f_4$. On the other hand  $f_4$ benefits from the 51 balls, hence in this case the Bayesian advantage implies that $f_4 \succ f_3$. The paradox appears because none of the reviewed models can represent this dual behavior. As a consequence, the Machina paradox, as well as the construction of a unified framework explaining both the Ellsberg and Machina paradoxes, are still open problems in decision making.
   
Ambiguity in economics is typically considered as a situation without a unique probability model describing it as opposed to \emph{risk}, which is defined as a situation with such a probability model describing it. It is however presupposed usually that a classical probability model is considered, defined on a $\sigma$--algebra of events. In the above approaches generalizing EUT \cite{gilboa,gilboaschmeidler1989,mmr2006,kmm2005}, more general structures are considered than that of a single classical probability model on a $\sigma$--algebra. Having looked in detail at the above mentioned structural generalizations, it can be noticed however that they all envisage generalizations of some specific aspects of the traditional situation of one classical probability model on one $\sigma$--algebra. 
Recently we have also proposed an approach to this problem, introducing the notion of {\it contextual risk}, inspired by the probability structure of quantum mechanics, which is intrinsically different from a classical probability on a $\sigma$-algebra, the set of events is indeed {\it not} a Boolean algebra \cite{aertsczachordhooghe2011,aertsdhooghesozzo2011,aertssozzovaxjo,aertssozzokavala12}.

In the present article we work out a direct mathematical representation of the Ellsberg and Machina paradox situations, in the standard formalism of quantum mechanics, hence by using a complex Hilbert space, and representing the probability measures by projection valued measures on this complex Hilbert space. As we will see when we explain the details of the Hilbert space representation of Ellsberg and Machina, it is not only the structure of the probability models which is essentially different from the known approaches -- projection valued measures instead of $\sigma$--algebra valued measures -- but also the way in which states are represented in quantum mechanics, i.e. by unit vectors of the Hilbert space, brings in an essential different aspect, coping both mathematically and intuitively, with the notion of ambiguity as introduced in economics.

\section{A quantum model for the Ellsberg paradox\label{ellsberg}}
To work out a quantum model for the Ellsberg paradox situation we consider the example resumed in Tab. 1, Sec. \ref{intro}. We will realize the quantum model in the three dimensional complex Hilbert space $\mathbb{C}^3$. Let us denote its canonical basis by the vectors $\{|1,0,0\rangle, |0,1,0\rangle, |0,0,1\rangle \}$. 

For the sake of clarity, we introduce the model in different steps. First, we define the part of the model, which we will refer to as the conceptual Ellsberg entity, as it consists of the Ellsberg situation without considering the different actions and also without considering the person and the bet to be taken. Hence it is the situation of the urn with 30 red balls and 60 black and yellow balls in unknown proportion. In the next steps we will add the remaining elements.

Already in the first part the presence of the ambiguity can be taken into account mathematically in a specific way by the quantum mechanical formalism. To this aim we introduce a quantum mechanical context
$e$, 
represented by means of the spectral family $\{P_r, P_{yb}\}$, where $P_r$ is the one dimensional orthogonal projection operator on the subspace generated by the vector $|1,0,0\rangle$, and $P_{yb}$ is the two dimensional orthogonal projection operator on the subspace generated by the vectors $|0,1,0\rangle$ and $|0,0,1\rangle$, $\{P_r, P_{yb}\}$ is indeed a spectral family, since $P_r \perp P_{yb}$ and $P_r+P_{yb}=\mathbbmss{1}$. Contexts, or more specifically measurement contexts, are indeed represented by spectral families of orthogonal projection operators, or by a self--adjoint operator determined by such a family. A state in quantum mechanics is represented by a unit vector of the complex Hilbert space. For example, the vector
\begin{equation} \label{stateredyellow}
|v_{ry}\rangle=|1/\sqrt{3}\cdot e^{i\theta_r}, \sqrt{2/3}\cdot e^{i\theta_y}, 0\rangle
\end{equation}
can be used to represent a state describing the Ellsberg situation. Indeed, we have
\begin{eqnarray}
|\langle 1,0,0|v_{ry}\rangle|^2=\langle v_{ry}|1,0,0\rangle\langle 1,0,0|v_{ry}\rangle=\parallel P_{r}|v_{ry}\rangle\parallel^{2}=1/3
\end{eqnarray}
which shows that the probability for `red' in the state represented by $|v_{ry}\rangle$ equals $1/3$. On the other hand, we have
\begin{equation}
\|P_{yb}|v_{ry}\rangle\|^{2}=\langle 0,\sqrt{2/3}\cdot e^{i\theta_y},0|0,\sqrt{2/3}\cdot e^{i\theta_y},0\rangle=2/3
\end{equation}
which shows that the probability for `yellow or black' in the state represented by $|v_{ry}\rangle$ is $2/3$. But this is not the only state describing the Ellsberg situation, the set of all such states (\emph{Ellsberg state set}) is
\begin{equation}
\Sigma_{Ells}=\{p_{v}:  \ |v\rangle=  |1/\sqrt{3}\cdot e^{i\theta_r},\rho_ye^{i\theta_y},\rho_be^{i\theta_b}\rangle \ \vert \ 0\le\rho_y,\rho_b,\ \rho_y^2+\rho_b^2=2/3\}
\end{equation}
which is a subset of $\mathbb{C}^3$. 
A state 
contained in $\Sigma_{Ells}$, 
together with the context $e$ represented by the spectral family $\{P_r, P_{yb}\}$, 
delivers a quantum description of the Ellsberg situation.

We come now to the second step, namely the introduction of a description of the different actions $f_1$, $f_2$, $f_3$ and $f_4$. Here a second measurement context is introduced which we denote by $g$. 
It describes the ball taken out of the urn, and its color verified, red, yellow or black. Also $g$ is represented by a spectral family of orthogonal projection operators $\{P_r, P_y, P_b\}$, where $P_r$ is already defined, while $P_y$ is the orthogonal projection operator on $|0,1,0\rangle$ and $P_b$ is the orthogonal projection operator on $|0, 0, 1\rangle$. This means that the probabilities, given a state $p_{v}$ represented by the vector $|v\rangle=|\rho_re^{i\theta_r},\rho_ye^{i\theta_y},\rho_be^{i\theta_b}\rangle$, are
\begin{eqnarray}
\mu_r(g,p_{v})=\parallel P_r|v\rangle\parallel^{2}=\langle v|P_{r}|v\rangle=\rho_r^2 \\
\mu_y(g,p_{v})=\parallel P_y| v\rangle\parallel^{2}=\langle v|P_{y}|v\rangle=\rho_y^2 \\
\mu_b(g,p_{v})=\parallel P_b|v\rangle\parallel^{2}=\langle v|P_{b}|v\rangle =\rho_b^2
\end{eqnarray}
where $\mu_r(g,p_{v})$, $\mu_y(g,p_{v})$ and $\mu_b(g,p_{v})$ are the probabilities to draw a red ball, a yellow ball and a black ball, respectively, in the state $p_{v}$. Of course,  if $p_v \in \Sigma_{Ells}$  we require that $\rho_r^{2}=1/3$ and $\rho_y^{2}+\rho_b^{2}=2/3$.

The different actions $f_1$, $f_2$, $f_3$ and $f_4$ are observables, and hence represented by self-adjoint operators, built all on the spectral decomposition $\{P_r, P_y, P_b\}$.
\begin{equation}
\hat f_1=12\$P_r \quad \hat f_2=12\$P_b \quad
\hat f_3=12\$P_r+12\$P_y \quad \hat f_4=12\$P_y+12\$P_b=12\$P_{yb}
\end{equation}
Let us analyze now the expected payoffs connected with the different acts, or the utility. Let us remark here that for reasons of simplicity, we identify the utility with the expected payoff, although of course in general the utility is a much more general variable. This implies that we are considering a risk neutral agent.
Hence, consider an arbitrary state $p_{v} \in \Sigma_{Ells}$ and the acts $f_1$ and $f_4$. We have
\begin{eqnarray}
U(f_1,g,p_{v})&=&\langle v|\hat f_1|v \rangle=12\$\langle v|P_r|v \rangle=12\$\cdot 1/3=4\$ \\
U(f_4,g,p_{v})&=&\langle v|\hat f_4|v \rangle=12\$\langle v|P_{yb}|v \rangle=12\$\cdot 2/3=8\$
\end{eqnarray} 
which shows that both these utilities are completely {\it independent} of the considered state of $\Sigma_{Ells}$. They are {\it ambiguity free}. Consider now the acts $f_2$ and $f_3$, and again an arbitrary state $p_{v} \in \Sigma_{Ells}$. We have
\begin{eqnarray}
U(f_2,g,p_{v})&=&\langle v|\hat f_2| v \rangle=12\$\langle v |P_b| v \rangle=12\$\mu_b(g,p_{v}) \\
U(f_3,g,p_{v})&=&\langle v|\hat f_3| v \rangle=12\$\langle v |(P_r+P_y)| v \rangle=12\$(\mu_r(g,p_{v})+\mu_y(g,p_{v}))
\end{eqnarray}
which shows that both utilities depend heavily on the state $p_{v}$, due to the ambiguity where the two acts are confronted with. 

Let us now consider some extreme cases to see explicitly the dependence on the state. Consider, e.g., the states $p_{v_{ry}}$, introduced in (\ref{stateredyellow}), and $p_{v_{rb}}$ represented by the vector $|v_{rb}\rangle= |1/\sqrt{3}\cdot e^{i\phi_r},0,\sqrt{2/3}\cdot e^{i\theta_b}\rangle$. These states give rise for the act $f_2$ to utilities
\begin{eqnarray}
U(f_2,g,p_{v_{ry}})&=&12\$\mu_b(g,p_{v_{ry}})=12\$\cdot0=0\$ \\
U(f_2,g,p_{v_{rb}})&=&12\$\mu_b(g,p_{v_{rb}})=12\$\cdot 2/3=8\$.
\end{eqnarray}
This shows that a state $p_{v_{rb}}$ exists within the realm of ambiguity, where the utility of act $f_2$ is greater than the utility of act $f_1$, and also a state $p_{v_{ry}}$ exists within the realm of ambiguity, where the utility of act $f_2$ is smaller than the utility of act $f_1$. If we look at act $f_3$, we find for the two considered extreme states the following utilities
\begin{eqnarray}
U(f_3,g,p_{v_{ry}})&=&12\$(\mu_r(g,p_{v_{ry}})+\mu_y(g,p_{v_{ry}}))=12\$(1/3+2/3)=12\$ \\
U(f_3,g,p_{v_{rb}})&=&12\$(\mu_r(g,p_{v_{rb}})+\mu_y(g,p_{v_{rb}}))=12\$(1/3+0)=4\$.
\end{eqnarray}
We are in a very similar situation, namely one of the states gives rise to a greater utility, while the other gives rise to a smaller utility than the independent one obtained in act $f_4$. 

We 
come 
finally to the third step, and 
take into account the presence of the ambiguity in a proper way. Relying on quantum mechanical modeling of situations that violate the Sure--Thing Principle, such as the Hawaii situation \cite{aerts2009}, we 
put forward the hypothesis that the two extreme states $p_{v_{ry}}$ and $p_{v_{rb}}$ play a role in the mind of the person that is asked to 
bet. Hence, it is a superposition state of these two states that will guide the decision of the person to bet. Let us construct a general superposition state
$p_{v_s}$ of these two states. Hence the vector $|v_s\rangle$ representing
$p_{v_s}$ can be written as follows
\begin{equation} \label{superpositionEllsberg}
|v_s\rangle=ae^{i\alpha}|v_{rb}\rangle+be^{i\beta}|v_{ry}\rangle
\end{equation}
where $a$, $b$, $\alpha$ and $\beta$ are chosen in such a way that $\langle v_s|v_s\rangle=1$, which means that $1=(ae^{-i\alpha}\langle v_{rb}|+be^{-i\beta}\langle v_{ry}|)(ae^{i\alpha}|v_{rb}\rangle+be^{i\beta}|v_{ry}\rangle)=a^2+b^2+2ab/3\cdot \cos(\beta-\alpha+\theta_r-\phi_r)$, from which follows that
\begin{equation}
\cos(\beta-\alpha+\theta_r-\phi_r)=3(1-a^2-b^2)/2ab
\end{equation}

Straightforward calculations show that the transition probabilities in the superposition state $p_{v_s}$ are given by
\begin{eqnarray}
|\langle 1,0,0|v_s\rangle|^2&=&1/3\cdot (3-2a^2-2b^2)=\mu_r(g,p_{v_s}) \\
|\langle 0,1,0|v_s\rangle|^2&=&2/3\cdot b^2=\mu_y(g,p_{v_s}) \\
|\langle 0,0,1|v_s\rangle|^2&=&2/3\cdot a^2=\mu_b(g,p_{v_s})
\end{eqnarray}
and that we can represent a general superposition state as
\begin{equation}
|v_s\rangle=1/\sqrt{3}\cdot |ae^{i(\alpha+\phi_r)}+be^{i(\beta+\theta_r)}, \sqrt{2}be^{i(\beta+\theta_y)}, \sqrt{2}ae^{i(\alpha+\theta_b)}\rangle
\end{equation}
and that the utilities corresponding to the observables of the different actions are given by
\begin{eqnarray}
U(f_1,g,p_{v_s})&=&\langle v_s|\hat f_1|v_s\rangle=12\$\cdot 1/3\cdot(3-2a^2-2b^2)=4\$\cdot(3-2a^2-2b^2) \\
U(f_2,g,p_{v_s})&=&\langle v_s|\hat f_2|v_s\rangle=12\$\cdot 2/3\cdot a^2=4\$\cdot 2a^2 \\
U(f_3,g,p_{v_s})&=&\langle v_s|\hat f_3|v_s\rangle=12\$\cdot 1/3\cdot(3-2a^2-2b^2)+12\$\cdot2/3\cdot b^2
                =4\$\cdot(3-2a^2) \\
U(f_4,g,p_{v_s})&=&\langle v_s|\hat f_4|v_s\rangle=12\$\cdot 2/3\cdot b^2+12\$\cdot2/3\cdot a^2=4\$(2a^2+2b^2) 
\end{eqnarray}
We can see that it is not necessarily the case that $\mu_r(g,p_{v_s})=1/3$, which means that choices of $a$ and $b$ can be made such that the superposition state $p_{v_s}$ is not a state contained in $\Sigma_{Ells}$. The reason is that $\Sigma_{Ells}$ is not a linearly closed subset of $\mathbb{C}^3$. 
A conservative choice within our quantum modeling is that we require the superposition state to be an element of $\Sigma_{Ells}$ -- we plan in future work to explore situations where this is not the case, e.g. situations of interference with respect to Ellsberg-type examples -- and this leads to
\begin{equation}
1/3=\mu_r(g,p_{v_s})=1/3\cdot (3-2a^2-2b^2) \Leftrightarrow a^2+b^2=1
\end{equation}
which implies that
$\cos(\beta-\alpha+\theta_r-\phi_r)=0$
and hence
$\beta=\pi/2+\alpha-\theta_r+\phi_r$.
Let us construct now two examples of superposition states that conserve the $1/3$ probability for drawing a red ball, and hence are conservative superpositions, and express the ambiguity as is thought to be the case in the Ellsberg paradox situation. The first state refers to the comparison for a bet between $f_1$ and $f_2$. The ambiguity of not knowing the number of yellow and black balls in the urn, only their sum to be 60, as compared to knowing the number of red balls in the urn to be 30, gives rise to the thought that `eventually there are perhaps almost no black balls and hence an abundance of yellow balls'. Jointly, and in superposition, the thought also comes that `it is of course also possible that there are more black balls than yellow balls'. These two thoughts in superposition, are mathematically represented by a state $p_{v_s}$. The state $p_{v_s}$ will be closer to $p_{v_{ry}}$, the extreme state with no black balls, if the person has a lot of ambiguity aversion, while it will be closer to $p_{v_{rb}}$, the extreme state with no yellow balls, if the person is attracted by the ambiguity. Hence, these two tendencies are expressed by the values of $a$ and $b$ in the superposition state. If we consider again the utilities, this time with $a^2+b^2=1$, we have
\begin{eqnarray}
U(f_1,g,p_{v_s})&=&4\$ \quad \quad \quad \quad \quad
U(f_2,g,p_{v_s})=4\$\cdot2a^2 \\
U(f_3,g,p_{v_s})&=&4\$\cdot(3-2a^2) \quad
U(f_4,g,p_{v_s})=8\$
\end{eqnarray}
So, for $a^2 < 1/2$, which exactly means that the superposition state $p_{v_s}$ is closer to the state $p_{v_{ry}}$ than to the state $p_{v_{rb}}$, we have that $U(f_2,g,p_{v_s}) < U(f_1,g,p_{v_s})$, and hence a person with strong ambiguity aversion in the situation of the first bet, will then prefer to bet on $f_1$ and not on $f_2$. Let us choose a concrete state for the bet between $f_1$ and $f_2$, and call it $p_{v_s^{12}}$, and denote its superposition state by $|v_s^{12}\rangle$. Hence, for $|v_s^{12}\rangle$ we take $a=1/2$ and $b=\sqrt{3}/2$ and hence $a^2=1/4$ and $b^2=3/4$. For the angles we must have $\beta-\alpha+\theta_r-\phi_r=\pi/2$, hence let us choose $\theta_r=\phi_r=0$, $\alpha=0$, and $\beta=\pi/2$. This gives us
\begin{equation}
|v_s^{12}\rangle=1/2\sqrt{3}\cdot |1+\sqrt{3}e^{i\pi/2}, \sqrt{2}\sqrt{3}e^{i\pi/2}, \sqrt{2}\rangle=1/2\sqrt{3}\cdot |1+i\sqrt{3}, i\sqrt{6}, \sqrt{2}\rangle
\end{equation}
On the other hand, for $1/2 < a^2$, which means that the superposition state is closer to the state $p_{v_{rb}}$ than to the state $p_{v_{ry}}$, we have that $U(f_3,g,p_{v_s}) < U(f_4,g,p_{v_s})$, and hence a person with strong ambiguity aversion in the situation of the second bet, will then prefer to bet on $f_4$ and not on $f_3$. Also for this case we construct an explicit state, let us call it $p_{v_s^{34}}$, and denote it by the vector $|v_s^{34}\rangle$. Hence, for $|v_s^{34}\rangle$ we take $a=\sqrt{3}/2$ and $b=1/2$ and hence $a^2=3/4$ and $b^2=1/4$. For the angles we must have $\beta-\alpha+\theta_r-\phi_r=\pi/2$, hence let us choose $\theta_r=\phi_r=0$, $\alpha=0$, and $\beta=\pi/2$. This gives us
\begin{equation}
|v_s^{34}\rangle=1/2\sqrt{3}\cdot |\sqrt{3}+e^{i\pi/2}, \sqrt{2}e^{i\pi/2}, \sqrt{2}\sqrt{3}\rangle=1/2\sqrt{3}\cdot |\sqrt{3}+i, i\sqrt{2}, \sqrt{6}\rangle
\end{equation}

\section{A quantum model for the Machina paradox\label{machina}}
In this section we elaborate a quantum model for the Machina paradox which is similar to the model constructed for the Ellsberg paradox. To this aim let us consider again the payoff matrix for the Machina situation in Tab. 2, Sec. \ref{intro}.

We consider the four dimensional complex Hilbert space $\mathbb{C}^4$ endowed with the canonical basis $\{ |1,0,0,0\rangle,$ $|0,1,0,0\rangle, |0,0,1,0\rangle, |0,0,0,1\rangle \}$. First, we describe the conceptual Machina entity, consisting of the Machina situation without the different actions and also without the person and the bet to be taken. Hence it is the situation of the urn with 50 balls of type 1 or type 2 and 51 balls of type 3 or type 4. This is described by the context $e$ represented by the spectral family $\{P_{12}, P_{34}\}$, where $P_{12}$ is the two dimensional orthogonal projection operator on the subspace generated by $\{ |1,0,0,0\rangle, $ $|0,1,0,0\rangle \}$, and $P_{34}$ is the two dimensional orthogonal projection operator on the subspace generated by $\{ |0,0,1,0\rangle, |0,0,0,1\rangle \}$. 

Then, as in Sec. \ref{ellsberg}, let us define the set of states of the Machina situation  
which we call the {\it Machina state set}
\begin{equation}
\Sigma_{Mach}=\{ p_{v}: |v\rangle=|v_1,v_2,v_3,v_4 \rangle \ \vert \  |v_1|^2+|v_2|^2=50/101, |v_3|^2+|v_4|^2=51/101 \}.
\end{equation}
If a state is contained in $\Sigma_{Mach}$, this state delivers a quantum description of the Machina situation, together with the measurement $e$ represented by the spectral family $\{P_{12}, P_{34}\}$ in $\mathbb{C}^4$.
 
We come now to the second step, namely the introduction of a description of the different actions $f_1$, $f_2$, $f_3$ and $f_4$, by means of the introduction of a second measurement context which we denote by g, and which describes how a ball is taken from the urn, and it is verified whether it is of type 1, 2, 3 or 4. Hence $g$ is represented by the spectral family $\{P_1, P_2, P_3, P_4\}$, where $P_1$, $P_2$, $P_3$ and $P_4$ are the orthogonal projection operators on the subspaces generated by $|1,0,0,0\rangle$, $|0,1,0,0\rangle$, $|0,0,1,0\rangle$ and $|0,0,0,1\rangle$, respectively. Thus, 
the probability $\mu_{j}(g,p_v)$ to draw a ball of type j, $j=1,2,3,4$, in the state $p_v$ represented by the vector $|v\rangle=|v_1,v_2,v_3,v_4\rangle$, is given by
\begin{eqnarray}
\mu_1(g,p_v)= \langle v |P_1| v \rangle=|v_1|^2, \quad \mu_2(g,p_v)=\langle v |P_2| v \rangle=|v_2|^2, \\
\mu_3(g,p_v)= \langle v |P_3| v \rangle=|v_3|^2, \quad \mu_4(g,p_v)=\langle v |P_4| v \rangle=|v_4|^2,
\end{eqnarray}
Let us then calculate the expected utilities associated with each of the feasible acts, proceeding as in the Ellsberg case. The acts $f_1$ to $f_4$ are observables, we represent them by self-adjoint operators built on the spectral family $ \{P_1, P_2, P_3, P_4\}$ in the following way: $
\hat{f_1}=\$202P_1+\$202P_2+\$101P_3+\$101P_4$, $\hat{f_2}=\$202P_1+\$101P_2+\$202P_3+\$101P_4$, $\hat{f_3}=\$303P_1+\$202P_2+\$101P_3+\$0P_4$ and $\hat{f_4}=\$303P_1+\$101P_2+\$202P_3+\$0P_4$.

Then we find
\begin{eqnarray}
U(f_1,p_v)&=&202 \cdot |v_1|^2+202 \cdot |v_2|^2+101 \cdot |v_3|^2+ 101 \cdot |v_4|^2=151 \\ 
U(f_2,p_v)&=&202 |v_1|^2 + 101 |v_2|^2 + 202 |v_3|^2+ 101|v_4|^2 \\
U(f_3,p_v)&=&303 |v_1|^2 + 202 |v_2|^2 + 101 |v_3|^2 \\
U(f_4,p_v)&=&303 |v_1|^2 + 101 |v_2|^2 + 202 |v_3|^2
\end{eqnarray} 
and see that only $U(f_1,p_v)$ is independent of $p_v$. Finally, as we did in the case of Ellsberg, let us calculate the utility for the three acts $f_2$, $f_3$ and $f_4$ for the Machina entity being in the extreme states $p_{v_{13}}$ and $p_{v_{24}}$ represented by the vectors 
\begin{eqnarray}
|v_{13}\rangle=|\sqrt{50/101}\cdot e^{i\theta_1}, 0, \sqrt{51/101}\cdot e^{i\theta_3}, 0\rangle \\
|v_{24}\rangle= \mid 0,\sqrt{50/101}\cdot e^{i\theta_2},0,\sqrt{51/101}\cdot e^{i\theta_4}\rangle
\end{eqnarray}
We have $U(f_2,p_{v_{13}})=202$, $U(f_3,p_{v_{13}})=201$, $U(f_4,p_{v_{13}})=252$, $U(f_2,p_{v_{24}})=101$, $U(f_3,p_{v_{24}})=100$ and $U(f_4,p_{v_{24}})=50$, which shows that for the state $p_{v_{13}}$ the utilities of all three acts $f_2$, $f_3$ and $f_4$ are maximal, and much bigger than the utility of $f_1$ as state independent act without ambiguity.
On the contrary, for the state $p_{v_{24}}$, we are in the inverse situation, for all three acts $f_2$, $f_3$ and $f_4$ the utilities are minimal, and much smaller than the utility of act $f_1$. Let us consider a superposition state $|v_s\rangle$ of these two extreme states
\begin{equation}
|v_s\rangle=ae^{i\alpha}|v_{13}\rangle+be^{i\beta}|v_{24}\rangle
\end{equation}
where $a$, $b$ are such that $a^2+b^2=1$, and $\alpha$ and $\beta$ are arbitrary, because indeed this makes $\langle v_s|v_s\rangle=1$, because $|v_{13}\rangle$ and $|v_{24}\rangle$ are orthogonal. We have
\begin{eqnarray}
\langle 1,0,0,0|v_s\rangle= ae^{i\alpha}\langle 1,0,0,0|v_{13}\rangle+be^{i\beta}\langle 1,0,0,0|v_{24}\rangle=a\sqrt{50/101}e^{i(\alpha+\theta_1)} \\
\langle 0,1,0,0|v_s\rangle= ae^{i\alpha}\langle 0,1,0,0|v_{13}\rangle+be^{i\beta}\langle 0,1,0,0|v_{24}\rangle=b\sqrt{50/101}e^{i(\beta+\theta_2)} \\
\langle 0,0,1,0|v_s\rangle= ae^{i\alpha}\langle 0,0,1,0|v_{13}\rangle+be^{i\beta}\langle 0,0,1,0|v_{24}\rangle=a\sqrt{51/101}e^{i(\alpha+\theta_3)} \\
\langle 0,0,0,1|v_s\rangle= ae^{i\alpha}\langle 0,0,0,1|v_{13}\rangle+be^{i\beta}\langle 0,0,0,1|v_{24}\rangle=b\sqrt{51/101}e^{i(\beta+\theta_4)}
\end{eqnarray}
which shows that
\begin{eqnarray}
|\langle 1,0,0,0|v_s\rangle|^2&=&50a^2/101=\mu_1(g,p_{v_s}) \\
|\langle 0,1,0,0|v_s\rangle|^2&=&50b^2/101=\mu_2(g,p_{v_s}) \\
|\langle 0,0,1,0|v_s\rangle|^2&=&51a^2/101=\mu_3(g,p_{v_s}) \\
|\langle 0,0,0,1|v_s\rangle|^2&=&51b^2/101=\mu_4(g,p_{v_s})
\end{eqnarray}
This means that for the utilities in this superposition state we find
\begin{eqnarray}
U(f_1,p_v)&=&202 \cdot 50a^2/101+202 \cdot 50b^2/101+101 \cdot 51a^2/101+ 101 \cdot 51b^2/101 \nonumber \\
&=&2\cdot 50(a^2+b^2)+51(a^2+b^2)=151 \\
U(f_2,p_v)&=&202 \cdot 50a^2/101 + 101 \cdot 50b^2/101 + 202 \cdot 51a^2/101+ 101 \cdot 51b^2/101 \nonumber \\
&=& 2\cdot 50 a^2+50b^2 +2\cdot 51a^2 + 51b^2 =202a^2+101b^2\\
U(f_3,p_v)&=&303 \cdot 50a^2/101 + 202 \cdot 50b^2/101 + 101 \cdot 51a^2/101 \nonumber \\
&=&3\cdot 50a^2+2\cdot 50b^2+51a^2=201a^2+100b^2 \\
U(f_4,p_v)&=&303 \cdot 50a^2/101 + 101 \cdot 50b^2/101 + 202 \cdot 51a^2/101 \nonumber \\
&=&3\cdot 50a^2+50b^2+2\cdot 51a^2=252a^2+50b^2
\end{eqnarray}
An important fact to mention before we further our quantum description, is that there is ample and convincing experimental evidence showing that ambiguity aversion is not  
related to the size of the payoffs involved \cite{camerer1999}. This means that if we want to model 
the effect of ambiguity, we should 
identify it mainly on the level of the states of the Ellsberg and Machina situations, and only on the level of the utilities as far as we take into account that it should not be linked to the size of the payoffs. 
This is exactly what we have done in our quantum model in the case of the Ellsberg paradox situation. Indeed, in equation (\ref{superpositionEllsberg}), we have considered a superposition state of the two extreme ambiguity states, and put forward the hypothesis that depending on the ambiguity aversion of a person, he or she will consider the Ellsberg conceptual entity in a state closer to one, or to the other, of the extreme states. Let us analyze the Machina situation in an analogous way now.

For example, consider the situation of a bet on $f_1$ or $f_2$. There is no ambiguity on $f_1$, since all states give rise to the same payoff, whereas there is a lot of ambiguity on $f_2$. A person with strong ambiguity aversion will consider the Machina conceptual entity to be in a superposition state close to the extreme state $p_{v_{24}}$, hence the value of $a$ will be small, and the value of $b$ large. Let us introduce the following values as an example to make a quantitative calculation possible, we take $a=1/\sqrt{10}$ and $b=3/\sqrt{10}$. Then we have $U(f_1,p_v)=151$, and $U(f_2,p_v)=202/10+101\cdot 9/10=20.2+90.9=111.1$. Hence, $U(f_2,p_v) < U(f_1,p_v)$, and this person will bet on $f_1$ and not on $f_2$. 

Consider now the situation of a bet on $f_3$ or $f_4$. In this case, for both actions there is an equal amount of ambiguity. This means that in principle no preference is present on the level of the `ambiguity choice' with respect to the superposition state that a person will consider the Machina conceptual entity to be in. Statistically this amounts to the superposition state being with equal values of $a$ and $b$ and hence we have $a=b=1/\sqrt{2}$. Let us calculate for these values of $a$ and $b$ the utilities corresponding to these actions. We have $U(f_3,p_v)=201/2+100/2=150.5$ and $U(f_4,p_v)=252/2+50/2=151$. This means that $U(f_3,p_v) < U(f_4,p_v)$ and hence the person will bet on $f_4$ and not on $f_3$.

We conclude with some remarks.

(i) The subjective preference in traditional economics approaches is incorporated in our approach in the quantum state, which represents what we have called the conceptual entity of the Ellsberg and Machina paradox situations. Hence, the subjective probabilities of the traditional economics approaches are captured by this quantum state, since it is introduced as describing the `conceptual entity', and not the 'physical entity'. At variance with existing proposals, the subjective preference, in our case incorporated in the quantum state, can be different for each one of the acts $f_j$, since it is not derived from the mathematical structure of the state space, or of the structure of other aspects of the Machina situation modeling. In the other approaches such a mathematical rule exists, which renders the Machina situation with $f_1$ preferred to $f_2$ and $f_4$ preferred to $f_3$ impossible. We have just seen that this is not impossible in our modeling scheme. 

(ii) Since in our approach there is no mathematical rule for the subjective probability measure that arises from the interaction of the person with the Machina conceptual entity, hence, from the interaction between the conceptual landscape carried by the person and the Machina conceptual entity, there is no problem to construct the exact probability measure, 
i.e. the exact 
superposition, that will model the experimental data, in case real experimental data are collected for the Machina example.

(iii) All existing proposals mathematically lead to a subjective probability, hence in our quantum model, to a specific superposition state, 
`as if this subjective probability 
i.e. this superposition state 
could be determined from a theoretical perspective'. We believe that the specific structure of this probability depends instead on the interaction of the betting person with the Ellsberg or Machina situation, and its values should be determined experimentally.

We have not yet explicitly introduced the quantum mechanical model of the bet itself. This is indeed another aspect of the quantum formalism where an essential deviation from the traditional economics approaches is bound to take place. Indeed, the bet itself, as a decision process, can be modeled within the same quantum mathematical formalism, by means of a spectral family of projection operators of a self-adjoint operator. We have done this explicitly already for the Ellsberg situation, and were able to model the experimental data of an experimental test of the Ellsberg paradox we published in \cite{aertsdhooghesozzo2011}. Due to space limitations of the present paper we decided to leave this part of the quantum mathematical model of Ellsberg and Machina for a forthcoming publication. 

In future work we also plan to investigate the relation to the already existing and fruitful approaches of introducing quantum structures in situations of decision under uncertainty in economics and decision theory \cite{busemeyerlambertmogiliansky2009}.

\end{document}